\documentclass[journal,10pt]{IEEEtran}

\usepackage{cite}
\usepackage{amsmath,amssymb}
\usepackage{graphicx,bm,color,hyperref}

\usepackage{diagbox}
\usepackage{caption}
\usepackage{arydshln}

\def\ket#1{|#1\rangle}
\def\bra#1{\langle#1|}
\def\braket#1#2{\langle#1|#2\rangle}
\def\C{\mathbb{C}}
\def\F{\mathbb{F}}

\def\wgt{\mathop{\rm wgt}\nolimits}

\newtheorem{theorem}{Theorem}

\newtheorem{lemma}[theorem]{Lemma}

\begin{document}

\title{Codes for Simultaneous Transmission of Quantum and Classical
  Information}

\author{\IEEEauthorblockN{Markus Grassl\IEEEauthorrefmark{1}, Sirui Lu\IEEEauthorrefmark{2}\IEEEauthorrefmark{3}, and Bei Zeng\IEEEauthorrefmark{3}\IEEEauthorrefmark{4}\IEEEauthorrefmark{5}}\\\medskip
  \IEEEauthorblockA{\small
    \IEEEauthorrefmark{1}Max-Planck-Institut f\"ur die Physik des Lichts, Staudtstra{\ss}e 2, 91058 Erlangen, Germany\\
    \IEEEauthorrefmark{2}Department of Physics, Tsinghua University, Beijing, 100084, China\\
    \IEEEauthorrefmark{3}Institute for Advanced Study, Tsinghua University, Beijing, 100084, China\\
    \IEEEauthorrefmark{4}Department of Mathematics \&  Statistics, University of Guelph, Guelph, ON, N1G 2W1, Canada\\
    \IEEEauthorrefmark{5}Institute for Quantum Computing, University of Waterloo, Waterloo,  Ontario, N2L 3G1, Canada\vspace*{-5ex}}
}

\maketitle

\begin{abstract}
  We consider the characterization as well as the construction of
  quantum codes that allow to transmit both quantum and classical
  information, which we refer to as `hybrid codes'.  We construct
  hybrid codes $[\![n,k{:}m,d]\!]_q$ with length $n$ and distance $d$,
  that simultaneously transmit $k$ qudits and $m$ symbols from a
  classical alphabet of size $q$.  Many good codes such as
  $[\![7,1{:}1,3]\!]_2$, $[\![9,2{:}2,3]\!]_2$,
  $[\![10,3{:}2,3]\!]_2$, $[\![11,4{:}2,3]\!]_2$,
  $[\![11,1{:}2,4]\!]_2$, $[\![13,1{:}4,4]\!]_2$,
  $[\![13,1{:}1,5]\!]_2$, $[\![14,1{:}2,5]\!]_2$,
  $[\![15,1{:}3,5]\!]_2$, $[\![19,9{:}1,4]\!]_2$,
  $[\![20,9{:}2,4]\!]_2$, $[\![21,9{:}3,4]\!]_2$,
  $[\![22,9{:}4,4]\!]_2$  have been found. All these codes have better parameters than
  hybrid codes obtained from the best known stabilizer quantum codes.
\end{abstract}

\section{Introduction}
The simultaneous transmission of both quantum and classical
information over a quantum channel was initially investigated in
\cite{DeSh05} from an information theoretic point of view, and
followed up by many others (see,
e.\,g. \cite{yard2005simultaneous,hsieh2010entanglement,hsieh2010trading}).
It was shown that there is an advantage to address the two tasks of
transmitting both quantum and classical information simultaneously,
compared to independent solutions.

For the finite length case, however, there are not many constructions
of error-correcting codes for simultaneous transmission of quantum and
classical information in the literature. In \cite{KHB08}, the authors
consider the problem in the context of so-called entanglement-assisted
codes, i.\,e., when sender and receiver share perfect entanglement.
The examples given in \cite{KHB08}, however, fail to demonstrate an
advantage in terms of the parameters of the resulting codes when
compared to, e.\,g., stabilizer quantum codes.

Here we study codes for simultaneous transmission of quantum and
classical information, which we refer to as `hybrid quantum codes' or
just `hybrid codes'.  Using the framework of stabilizer codes
\cite{gottesman1997stabilizer,CRSS98} and its generalization, that is,
codeword stabilized (CWS) codes \cite{CSSZ09} and union stabilizer
codes \cite{GrRo08}, we obtain hybrid codes for up to eleven qubits by
exhaustive or randomized search. We have found many good hybrid codes
that have advantage over the best known quantum codes for transmitting
quantum information only.  A general construction yields codes for up
to $38$ qubits.  We also formulate a linear program to bound the
parameters of hybrid codes.

\section{Background and Notation}
Our discussion is based on the theory of stabilizer quantum codes and
its connection to classical error-correcting codes (see, e.\,g.,
\cite{CRSS98}).  Although we consider only codes for qubit systems
here, we state the theory for quantum systems composed of qudits of
dimension $q=p^\ell$, where $p$ is prime.  A quantum error-correcting
code, denoted by $\mathcal{C}=(\!(n,K,d)\!)_q$, is a $K$-dimensional
subspace of the Hilbert space $\mathcal{H}=(\C^q)^{\otimes n}$, which
is an $n$-fold tensor product of Hilbert spaces of dimension $q$.  If
the minimum distance of the code is $d$, then any error affecting no
more than $d-1$ of the subsystems can be detected or acts as a
multiple of identity on the code.  For stabilizer codes, the dimension
$K$ is a power of $p$, and if $K=q^k$, we use the notation
$\mathcal{C}=[\![n,k,d]\!]_q$.  For classical block codes, the
notation $C=(n,M,d)_q$ is used, and if the code is linear with
cardinality $M=q^m$, we use the notation $C=[n,m,d]_q$.  Following
\cite{KHB08}, we use the notation $\mathcal{C}=[\![n,k{:}m,d]\!]_q$ for
a code that simultaneously transmits $k$ qudits and $m$ symbols from a
classical alphabet of size $q$.  Similarly, we use the notation
$\mathcal{C}=(\!(n,K{:}M,d)\!)_q$ for such a code that encodes a quantum
system of dimension $K$ and one out of $M$ classical messages.

Trivially, we have the following:
\begin{lemma}\label{lemma:hybrid_from_QECC}
  Given a quantum code $\mathcal{C}=(\!(n,KM,d)\!)_q$ of composite
  dimension $KM$, there exists a hybrid code with parameters
  $(\!(n,K{:}M,d)\!)_q$.
\end{lemma}
\begin{IEEEproof}
  First, factor the code space into two subsystems of dimension $K$
  and $M$, respectively. Then, one uses the first subsystem of
  dimension $K$ to transmit quantum information, and the second
  subsystem of dimension $M$ just to transmit classical information.
\end{IEEEproof}
Similarly, we have the following conversion rule for hybrid stabilizer
codes.
\begin{lemma}\label{lemma:quantum_to_classical}
  Assume that a hybrid code $\mathcal{C}=[\![n,k{:}m,d]\!]_q$ with $k>0$
  exists. Then a code $\mathcal{C}'=[\![n,k-1{:}m+1,d]\!]_q$ exists as well.
\end{lemma}
\begin{IEEEproof}
  One of the $k$ qudits can be used to transmit classical information
  only, decreasing $k$ and increasing $m$.
\end{IEEEproof}
Note that the converse does not hold in general, as the transmission
of quantum information over a quantum channel is more demanding than
the transmission of classical information.

Another trivial construction is to independently use a quantum code of
length $n_1$ and a classical code of length $n_2$.
\begin{lemma}\label{lemma:independent_hybrid_codes}
  Assume that a quantum code $\mathcal{C}_1=[\![n_1,k_1,d]\!]_q$ and a
  classical code $C_2=[n_2,m_2,d]_q$ exist. Then there exists a hybrid
  code with parameters $\mathcal{C}=[\![n_1+n_2,k_1{:}m_2,d]\!]_q$.
\end{lemma}
Our goal is to find codes that have better parameters than the codes
that can be obtained by these trivial constructions.

\section{Error correction conditions}\label{sec:code_conditions}
A hybrid quantum code
$\mathcal{C}=(\!(n,K{:}M)\!)_q$ can be described by a collection
\begin{alignat}{5}
  \{ \mathcal{C}^{(\nu)}\colon\nu=1,\ldots,M\}
\end{alignat}
of $M$ quantum codes $\mathcal{C}^{(\nu)}=(\!(n,K,d)\!)_q$.  Each of
the codes has length $n$, dimension $K$, and minimum distance $d$.
The classical information $\nu$ determines which quantum code
$\mathcal{C}^{(\nu)}$ is used to encode the quantum information.
In the following, we will use Greek letters when referring to
classical information.
Assume that
\begin{alignat}{5}
  \{\ket{c_i^{(\nu)}}\colon i=1,\ldots,K\}
\end{alignat}
is an orthonormal basis for the code $\mathcal{C}^{(\nu)}$.  In order
to be able to correct the linear span of error operators $\{E_k\colon
k=1,2,\ldots\}$, each of the codes $\mathcal{C}^{(\nu)}$ has to obey
the Knill-Laflamme conditions \cite{KnLa97}, i.\,e.,
\begin{alignat}{5}
  \bra{c_i^{(\nu)}}E_k^\dagger E_\ell\ket{c_j^{(\nu)}}=\alpha_{k\ell}^{(\nu)}\delta_{ij}.\label{eq:cond_quantum}
\end{alignat}
Note that the constants $\alpha_{k\ell}^{(\nu)}\in\C$ may depend on the
classical information $\nu$. 

On the other hand, in order to be able to retrieve the classical
information $\nu$ independently of the quantum information that is
transmitted at the same time, one has to be able to perfectly
distinguish the states $\ket{c_i^{(\nu)}}$ and $\ket{c_j^{(\mu)}}$ for
$\nu\ne\mu$ and arbitrary $i$ and $j$ after an error.  This is
reflected by the condition
\begin{alignat}{5}
  \bra{c_i^{(\nu)}}E_k^\dagger E_\ell\ket{c_j^{(\mu)}}=0,\qquad\text{for $\mu\ne\nu$.}\label{eq:cond_classical}
\end{alignat}
In particular, the states $\ket{c_i^{(\nu)}}$ and $\ket{c_j^{(\mu)}}$
have to be mutually orthogonal.  Combining \eqref{eq:cond_quantum} and
\eqref{eq:cond_classical}, we get the following necessary and
sufficient condition for hybrid quantum codes.
\begin{theorem}
  A hybrid quantum code $\mathcal{C}=(\!(n,K{:}M)\!)_q$ with orthonormal
  basis states $\{\ket{c_i^{(\nu)}}\colon
  i=1,\ldots,K,\;\nu=1,\ldots,M\}$ can correct all errors $\{E_k\colon
  k=1,2,\ldots\}$ if and only if
  \begin{alignat}{5}
  \bra{c_i^{(\nu)}}E_k^\dagger E_\ell\ket{c_j^{(\mu)}}=\alpha_{k\ell}^{(\nu)}\delta_{ij}\delta_{\mu\nu}.\label{eq:cond_hybrid}
  \end{alignat}
\end{theorem}
\begin{IEEEproof}
  As argued above, for $\mu=\nu$ condition \eqref{eq:cond_hybrid}
  reduced to the Knill-Laflamme conditions. Now assume that
  $\mu\ne\nu$. When condition \eqref{eq:cond_classical} is violated,
  i.\,e., $\bra{c_i^{(\nu)}}E_k^\dagger E_\ell\ket{c_j^{(\mu)}}\ne 0$,
  the erroneous states $E_k\ket{c_i^{(\nu)}}$ and
  $E_\ell\ket{c_j^{(\mu)}}$ are non-orthogonal and can not be perfectly
  distinguished. On the other hand, when condition
  \eqref{eq:cond_classical} holds, then the spaces
  $\mathcal{V}^{(\nu)}$ spanned by the images of the code
  $\mathcal{C}^{(\nu)}$ under all error operators, i.\,e.,
  \begin{alignat}{5}
    \mathcal{V}^{(\nu)}=\left\langle E_k\ket{c_i^{(\nu)}}\colon i=1,\ldots K,\;k=1,2,\ldots\right\rangle\label{eq:error_spaces}
  \end{alignat}
  are mutually orthogonal. Therefore, there exists a measurement with
  associated orthogonal projections $P^{(\nu)}$ that can be used to
  retrieve the classical information $\nu$.  Then, knowing the index
  $\nu$, one can apply the decoding algorithm for the code
  $\mathcal{C}^{(\nu)}$ to retrieve the quantum information.
\end{IEEEproof}

Note that in the special case that the constants $\alpha_{k\ell}^{(\nu)}$
do not depend on $\nu$, condition \eqref{eq:cond_hybrid} reduces to
the Knill-Laflamme conditions for a quantum code
$\mathcal{C}=(\!(n,KM)\!)_q$ of dimension $KM$ with basis states
$\{\ket{c_i^{(\nu)}}\colon
i=1,\ldots,K,\;\nu=1,\ldots,M\}$. Therefore, for hybrid codes to have
better parameters than the codes given by Lemma
\ref{lemma:hybrid_from_QECC}, there should be at least a pair
$\nu,\mu$ and errors $E_k,E_\ell$ such that
$\alpha_{k\ell}^{(\nu)}\ne\alpha_{k\ell}^{(\mu)}$.  In particular, when the
error operators $E_k$ are unitary, $\alpha_{kk}^{(\nu)}=1$.  Then one
should have $\alpha_{k\ell}^{(\nu)}\ne 0$ for some $\nu$ and $k\ne\ell$,
which suggests that some of the codes $\mathcal{C}^{(\nu)}$ might be
taken to be degenerate codes.  In that case, the dimension of the
spaces $\mathcal{V}^{(\nu)}$ in \eqref{eq:error_spaces} is smaller,
and hence one might be able to find a larger number of such spaces
that are mutually orthogonal.  In general, however, it is not excluded
that all the subcodes $\mathcal{C}^{(\nu)}$ of a hybrid quantum code
$\mathcal{C}=(\!(n,K{:}M,d)\!)_q$ are non-degenerate and at the same
time the product $KM$ is strictly larger than the maximal dimension
$K'$ of any quantum code $\mathcal{C}'=(\!(n,K',d)\!)_q$.

An alternative characterization of hybrid quantum codes in the
Heisenberg picture of quantum mechanics was given as a special case in
\cite{BKK07}.

\section{Code construction}\label{sec:con}
We outline the construction of hybrid quantum codes in the framework
of CWS codes/union stabilizer codes.  We start with a quantum code
$\mathcal{C}^{(0)}=(\!(n,K,d)\!)_q$ which is a CWS code that might
even be a stabilizer code $\mathcal{C}^{(0)}=[\![n,k,d]\!]_q$.  The
codes $\mathcal{C}^{(\nu)}$ are chosen as images of the seed code
$\mathcal{C}^{(0)}$ under tensor products of generalized Pauli
matrices, denoted by $t_\nu$.  Thus we have
\begin{alignat}{5}
\mathcal{C}^{(\nu)}= t_\nu \mathcal{C}^{(0)}
\end{alignat}
with $\{t_\nu\colon\nu=1,\ldots M\}$ a set of $M$ translation
operators.  When $\mathcal{C}^{(0)}$ is a non-degenerate quantum code,
then all the codes $\mathcal{C}^{(\nu)}$ will also be non-degenerate.
Furthermore, in this situation $\alpha_{k\ell}^{(\nu)}=\delta_{k\ell}$ for
generalized Pauli errors $E_k$ and $E_\ell$.  Then the resulting code
will be a quantum code of dimension $KM$.  Therefore, the seed code
$\mathcal{C}^{(0)}$ is chosen to be degenerate.

Next we consider the classical codes associated with the quantum codes
$\mathcal{C}^{(\nu)}$.  For simplicity, we first consider the special
case of stabilizer codes.  The stabilizer group $\mathcal{S}$ of the
code $\mathcal{C}^{(0)}$ corresponds to a self-orthogonal classical
code $C_0$.  The code $C_0$ is contained in its symplectic dual
$C_0^*$, i.\,e., $C_0\subseteq C_0^*$, which corresponds to the
normalizer $\mathcal{N}$ of the stabilizer groups $\mathcal{S}$ in the
generalized $n$-qudit Pauli group.  For impure codes, we have
\begin{alignat}{5}
  d=\min\{\wgt{c}\colon c\in C_0^*\setminus C_0\}>\min\{\wgt{c}\colon c\in C_0^*\setminus\{0\}\}.
\end{alignat}
The codes $\mathcal{C}^{(\nu)}= t_\nu \mathcal{C}^{(0)}$ are
associated with cosets $C_0^*+t_\nu$ of the normalizer code $C_0^*$,
where we use the same symbol $t_\nu$ to denote the classical vector
corresponding to the translation operator.  When the cosets
$C_0^*+t_\nu$ and $C_0^*+t_\mu$ are different, then the codes
$\mathcal{C}^{(\nu)}$ and $\mathcal{C}^{(\mu)}$ will be orthogonal to
each other.  The hybrid quantum code $\mathcal{C}$ is associated with
the classical code
\begin{alignat}{5}
  C^*=\bigcup_{\nu=1}^M C_0^*+t_\nu.\label{eq:union_normalizer}
\end{alignat}
When the union of the codes in \eqref{eq:union_normalizer} is an
additive code, the hybrid quantum code will be a stabilizer code.
Note that, in general, we have the chain of classical codes
\begin{alignat}{5}
C \le C_0 \le C_0^*\le C^*.
\end{alignat}
The minimum distance of the quantum code associated with $C^*$ is
computed as
\begin{alignat}{5}
  d'=\min\{\wgt{c}\colon c\in C^*\setminus C\}.\label{eq:dmin_quantum}
\end{alignat}
It turns out that the minimum distance of a hybrid code associated
with the codes $C_0\le C^*$ is given by 
\begin{alignat}{5}
  d=\min\{\wgt{c}\colon c\in C^*\setminus C_0\}.\label{eq:dmin_hybrid}
\end{alignat}
Note that the minimum in \eqref{eq:dmin_hybrid} is taken over a
smaller set compared to \eqref{eq:dmin_quantum}, as $C\le C_0$, and
hence $d\ge d'$.

In summary, we have the following construction.
\begin{theorem}
  Let $C_0=(n,q^{n-k},d_0)_{q^2}$ be a classical additive code that is
  contained in its symplectic dual $C_0^*$.  Further, let
  $C^*=(n,q^{n+k+m},d')_{q^2}$ be an additive code containing
  $C_0^*$. Then there exists a hybrid stabilizer code
  $\mathcal{C}=[\![n,k{:}m,d]\!]_q$ encoding $k$ qudits and $m$
  classical symbols.  The minimum distance of $\mathcal{C}$ is given
  by
  \begin{alignat}{5}
    d=\min\{\wgt{c}\colon c\in C^*\setminus C_0\}.\label{eq:dmin_hybrid+theorem}
  \end{alignat}
\end{theorem}
\begin{IEEEproof}
  There are $q^m$ cosets of the code $C_0^*$ in the code $C^*$.  Using
  the representatives $t_\nu$ of the cosets $C^*/C_0^*$, we obtain the
  translated codes $\mathcal{C}^{(\nu)}=t_\nu\mathcal{C}^{(0)}$ which
  are mutually orthogonal.  All these codes have the same minimum
  distance given by
  \begin{alignat}{5}
    d''&{}=\min\{\wgt{c}\colon c\in C_0^*\setminus C_0\}\\
       &{}\ge \min\{\wgt{c}\colon c\in C^*\setminus C_0\}=d.
  \end{alignat}
  Hence, condition \eqref{eq:cond_hybrid} holds for $\nu=\mu$.  It
  remains to show that the distance between the quantum codes
  $C^{(\nu)}$ is at least $d$, i.\,e., that \eqref{eq:cond_classical}
  holds for all operators $E_k^\dagger E_\ell$ of weight at most
  $d-1$.  When we treat the linear span of all codes
  $\mathcal{C}^{(\nu)}$ as a larger stabilizer code, the minimum
  distance would be given by \eqref{eq:dmin_quantum}.  When
  $E_k^\dagger E_\ell$ is an element of the stabilizer of
  $\mathcal{C}^{(0)}$, for $\nu\ne\mu$ we compute
  \begin{alignat}{5}
      \bra{c_i^{(\nu)}}E_k^\dagger E_\ell\ket{c_j^{(\mu)}}
      &{}=\bra{c_i^{(0)}}t_\nu^\dagger E_k^\dagger E_\ell t_\mu\ket{c_j^{(0)}}\\
      &{}\propto\bra{c_i^{(0)}}t_\nu^\dagger t_\mu E_k^\dagger E_\ell\ket{c_j^{(0)}}\\
      &{}=\bra{c_i^{(0)}}t_\nu^\dagger t_\mu\ket{c_j^{(0)}}
      =\braket{c_i^{(\nu)}}{c_j^{(\mu)}}=0.
  \end{alignat}
  Hence we can not only exclude the elements of $C$, but also those of
  $C_0$ when computing the minimum distance in
  \eqref{eq:dmin_hybrid+theorem}.
\end{IEEEproof}
In terms of classical codes, the task of constructing a good hybrid
stabilizer code can be carried out in two steps.  First, one has to
find a good additive code $C_0^*$ that contains its symplectic dual
$C_0$.  This defines the seed code $\mathcal{C}^{(0)}$ used to encode
the quantum information.  Then, using $m$ additional generators for
encoding the classical information, one obtains the code $C$ with
$C_0^*\le C^*$.

\begin{table*}[!htb]
    \caption{(LP bound) Upper bound on the number of classical bits $m$
      in any $[\![n, k{:}m, d]\!]_2$ hybrid stabilizer code with fixed
      length $n\le 14$ and dimension $k$ for distance $d=3,4,5$. For $k=0$,
      we list the largest dimension of a classical linear binary
      code. Note that there is, e.\,g., no stabilizer code
      $[\![13,5,4]\!]_2$, excluding the corresponding entry in the
      table marked with ${}^*$.}
    \vskip2ex
    \hfill\begin{tabular}{c||c|c|c|c|c|c|c|c|c}
      \multicolumn{10}{c}{$d=3$}\\[2ex]
\diagbox{$n$}{$k$} & 0  & 1  & 2 & 3 & 4 & 5 & 6 & 7 & 8 \\ \hline\hline
5                  & 2  & 0  &-- &-- &-- &-- &-- &-- &-- \\
6                  & 3  & 0  &-- &-- &-- &-- &-- &-- &-- \\
7                  & 4  & 2  &-- &-- &-- &-- &-- &-- &-- \\
8                  & 4  & 3  & 1 & 0 &-- &-- &-- &-- &-- \\
9                  & 5  & 4  & 3 & 1 &-- &-- &-- &-- &-- \\
10                 & 6  & 5  & 4 & 2 & 1 &-- &-- &-- &-- \\
11                 & 7  & 6  & 5 & 4 & 2 & 0 &-- &-- &-- \\
12                 & 8  & 7  & 6 & 5 & 3 & 2 & 0 &-- &-- \\
13                 & 9  & 8  & 7 & 5 & 5 & 3 & 1 & 0 &-- \\
14                 & 10 & 9  & 8 & 7 & 6 & 5 & 3 & 1 & 0 
\end{tabular}\hfill\def\st{\rlap{${}^*$}}
    \begin{tabular}{c||c|c|c|c|c|c|c}
            \multicolumn{8}{c}{$d=4$}\\[2ex]
\diagbox{$n$}{$k$} & 0  & 1 & 2 & 3 & 4 & 5 & 6 \\ \hline\hline
5                  & 1  &-- &-- &-- &-- &-- &-- \\
6                  & 2  &-- &-- &-- &-- &-- &-- \\
7                  & 3  &-- &-- &-- &-- &-- &-- \\
8                  & 4  &-- &-- &-- &-- &-- &-- \\
9                  & 4  &-- &-- &-- &-- &-- &-- \\
10                 & 5  & 3 & 1 &-- &-- &-- &-- \\
11                 & 6  & 4 & 2 &-- &-- &-- &-- \\
12                 & 7  & 5 & 4 & 2 & 0 &-- &-- \\
13                 & 8  & 6 & 5 & 4 & 2 & 0\st &-- \\
14                 & 9  & 6 & 6 & 5 & 3 & 2 & 0 
\end{tabular}\hfill
    \begin{tabular}{c||c|c|c|c}
      \multicolumn{5}{c}{$d=5$}\\[2ex]
\diagbox{$n$}{$k$} & 0 & 1 & 2 & 3  \\ \hline\hline
5                  & 1 &-- &-- &-- \\
6                  & 1 &-- &-- &-- \\
7                  & 1 &-- &-- &-- \\
8                  & 2 &-- &-- &-- \\
9                  & 2 &-- &-- &-- \\
10                 & 3 &-- &-- &-- \\
11                 & 4 & 0 &-- &-- \\
12                 & 4 & 2 &-- &-- \\
13                 & 5 & 4 &-- &-- \\
14                 & 6 & 5 & 3 & 1 
\end{tabular}\hfill    
    \label{tab:LP_bounds}
    \vskip2ex
\end{table*}

\section{Linear programming bounds}
In order to obtain bounds on the parameters of hybrid stabilizer codes
$[\![n,k{:}m,d]\!]_q$, we consider the homogeneous weight enumerators of
the associated code $C_0$ and its symplectic dual $C_0^*$, as well as
the code $C^*$  and its symplectic dual $C$:
\begin{alignat}{5}
  \mathcal{W}_{C_0}(X,Y)&{}=\sum_{w=0}^n A_w^\bot X^{n-w}Y^w,\\
 \mathcal{W}_{C_0^*}(X,Y)&{}=\sum_{w=0}^n A_w X^{n-w}Y^w,\\
  \mathcal{W}_{C}(X,Y)&{}=\sum_{w=0}^n B_w^\bot X^{n-w}Y^w,\\
 \mathcal{W}_{C^*}(X,Y)&{}=\sum_{w=0}^n B_w X^{n-w}Y^w.
\end{alignat}
The weight enumerators of $C_0$ and $C_0^*$, as well as those of $C$
and $C^*$, are related by the MacWilliams transformation, i.\,e.,
\begin{alignat}{5}
  \mathcal{W}_{C_0^*}(X,Y)&{}=\frac{1}{|C_0|}\mathcal{W}_{C_0}\left(X+(q^2-1)Y,X-Y\right),\label{eq:MacWilliams1}\\
  \mathcal{W}_{C^*}(X,Y)&{}=\frac{1}{|C|}\mathcal{W}_{C}\left(X+(q^2-1)Y,X-Y\right).\label{eq:MacWilliams2}
\end{alignat}
Nestedness of the codes implies the condition
\begin{alignat}{5}
  0\le B_w^\bot\le A_w^\bot \le A_w \le B_w,\qquad\text{for $w=0,\ldots,n$.}\label{eq:WE_nested_codes}
\end{alignat}
When the hybrid code has minimum distance $d$, we have
\begin{alignat}{5}
  A_w^\bot= A_w = B_w,\qquad\text{for $w=0,\ldots,d-1$.}\label{eq:WE_dmin}
\end{alignat}
Additionally, we have:
\begin{alignat}{9}
  A_0^\bot = A_0 &{}= B_0 =1,\\
  \sum_{w=0}^n A_w^\bot &{}= q^{n-k},\quad&
  \sum_{w=0}^n A_w &{}= q^{n+k},\\
  \sum_{w=0}^n B_w^\bot &{}= q^{n-k-m},\quad&
  \sum_{w=0}^n B_w &{}= q^{n+k+m}.\label{eq:sum_B_w}
\end{alignat}
When a hybrid stabilizer code $[\![n,k{:}m,d]\!]_q$ exists, the linear
program for the variables $B_w^\bot$, $A_w^\bot$, $A_w$, and $B_w$
given by \eqref{eq:MacWilliams1}--\eqref{eq:sum_B_w} has an integer
solution.  For qubit codes, we can strengthen the linear program by
additionally considering the shadow enumerator \cite{Rains99}
\begin{alignat}{5}
  \mathcal{S}_{C_0}(X,Y)&{}=\frac{1}{|C_0|}\mathcal{W}_{C_0}\left(X+(q^2-1)Y,Y-X\right),\label{eq:shadow}
\end{alignat}
which has to have non-negative integer coefficients.

Using CPLEX V12.6.3.0, we checked whether the integer program is
feasible.  More precisely, we first fix the length $n$, number of
qudits $k$, and number $M=2^m$ of classical symbols. Then we look for
the largest minimum distance $d$ for which the integer program is
found to be feasible.  The resulting bounds on the parameters
$[\![n,k{:}m,d]\!]_2$ are listed in Table \ref{tab:LP_bounds}, i.\,e.,
for fixed parameters $n$, $k$, and $d$, the largest possible value for
$m$ is given.  For $n>14$, there seem to be some precision issues, so
we list only the bounds for $n\le 14$.

\section{Results}
Based on the construction discussed in Section~\ref{sec:con}, we
perform a search for $\mathcal{C}=[\![n,k{:}m,d]\!]_2$ codes with
distance $d\ge 3$.  We start with the self-dual codes from the
classification in \cite{DaPa06,Danielsen_web}.  In a first step, we
construct impure quantum codes $[\![n,1,d]\!]_2$, and then look for
additional vectors for the encoding of classical information,
resulting in an $[\![n,1{:}m',d]\!]_2$ hybrid code.  In some cases it
turns out that we can encode more than one qubit, i.\,e., the code
$[\![n,1{:}m',d]\!]_2$ is in fact a hybrid code with parameters
$[\![n,k{:}m'-k+1,d]\!]_2$.

For distance $d=4$ and $d=5$, we have exhaustively searched using all
self-dual codes listed in \cite{DaPa06,Danielsen_web} up to length
$n=11$. For $d=3$, we have exhaustively searched all self-dual codes
listed in \cite{DaPa06,Danielsen_web} up to $n=10$.  We also have
conducted randomized search for $n=11$. Finally, we appended some
qubits in the state $\ket{0}$ to good quantum codes and found new
hybrid codes.  The results are summarized as follows.
\begin{theorem}
  There exist hybrid codes with the following parameters:
  \begin{alignat*}{10}
  & [\![7,1{:}1,3]\!]_2,  &\:& [\![9,2{:}2,3]\!]_2,  &\:& [\![10,3{:}2,3]\!]_2, &\:& [\![11,4{:}2,3]\!]_2,\\
  & [\![11,1{:}2,4]\!]_2, &\:& [\![13,1{:}4,4]\!]_2, &\:&\\
  & [\![13,1{:}1,5]\!]_2, &\:& [\![14,1{:}2,5]\!]_2, &\:& [\![15,1{:}3,5]\!]_2,\\
  & [\![19,9{:}1,4]\!]_2, &\:& [\![20,9{:}2,4]\!]_2, &\:& [\![21,9{:}3,4]\!]_2, &\:& [\![22,9{:}4,4]\!]_2.
  \end{alignat*}
  All these codes have better parameters than codes obtained from the
  best quantum codes using Lemma \ref{lemma:quantum_to_classical}.
\end{theorem}

Below, we provide more details on these codes.  In presenting each
$[\![n,k{:}m,d]\!]_2$ code, we first list the generators of the
stabilizer of the corresponding $[\![n,k,d]\!]_2$ impure quantum code
$\mathcal{C}^{(0)}$, with its $2k$ logical operators between a single
and a double horizontal line.  The stabilizer of the code
$\mathcal{C}^{(0)}$, corresponding to the classical code $C_0$, is
generated by the rows above the single horizontal line, while the
normalizer of the code $\mathcal{C}^{(0)}$, corresponding to the
symplectic dual code $C_0^*$, is generated by the rows above the
double horizontal line.  Below the double horizontal line, we list the
additional generators that are used to encode $m$ classical bits.

A quantum code that encodes a single qubit and is able to correct a
single error requires at least five qubits.  For five and six qubits,
linear programming shows that we can only transmit a single qubit and
no additional classical bit when we want to correct a single errors,
i.\,e., for distance $d\ge 3$.

Increasing the length to seven qubits, it is still only possible to
encode a single qubit when a single error has to be corrected.  The
stabilizer of an impure code $[\![7,1,3]\!]_2$ is generated by the
elements of the Pauli group given in first six lines above the single
horizontal line in the matrix \eqref{eq:gen_7_1_1_3}.  Note that the
element in the second line has only weight two.  The next two elements
between the single and the double horizontal line correspond to the
logical operators on the encoded qubit.  Starting with this impure
code, we are able to transmit an extra classical bit, i.\,e., we
obtain a hybrid code with parameters $[\![7,1{:}1,3]\!]_2$. The
additional generator that is used to encode one classical bit is given
below the double horizontal line.
\begin{table*}
    \caption{Generators of hybrid codes $[\![9,2{:}2,3]\!]_2$,
      $[\![10,3{:}2,3]\!]_2$, and $[\![11,1{:}2,4]\!]_2$. }
    \hfill\abovedisplayskip0pt
    \begin{minipage}{0.3\textwidth}
  \begin{footnotesize}
  \begin{alignat}{5}\arraycolsep0.5\arraycolsep
    \left(\begin{array}{*{9}{c}}
      X & I & I & Z & Y & Z & X & X & Y\\
      Z & I & I & I & I & X & I & I & I\\
      I & X & I & Z & Y & I & Y & I & Z\\
      I & Z & I & I & I & I & X & I & I\\
      I & I & X & Z & Z & I & I & I & X\\
      I & I & Z & I & Y & X & I & Y & I\\
      I & I & I & X & X & X & I & Z & I\\
      \hline
      I & I & I & Z & I & I & X & Y & X\\
      I & I & I & I & X & I & I & Z & Y\\
      I & I & I & I & Z & I & I & X & X\\
      I & I & I & I & I & X & X & I & X\\
      \hline\hline
      I & I & I & I & I & Z & I & Z & X\\
      I & I & I & I & I & I & Y & X & Z\\
    \end{array}\right)\label{eq;gen_9_2_2_3}
  \end{alignat}
  \end{footnotesize}
    \end{minipage}\hfill
    \begin{minipage}{0.33\textwidth}
    \begin{footnotesize}
  \begin{alignat}{5}\arraycolsep0.5\arraycolsep
  \left(\begin{array}{*{10}{c}}
    X & I & X & Y & I & X & Z & X & X & Y\\
    Z & I & I & I & I & I & I & I & I & X\\
    I & X & X & X & I & Y & X & Y & Z & X\\
    I & Z & I & I & I & I & I & I & X & I\\
    I & I & Z & Z & I & I & I & I & I & I\\
    I & I & I & I & X & X & Y & Y & I & I\\
    I & I & I & I & Z & Z & X & X & I & I\\
    \hline
    I & I & X & X & I & I & I & I & I & X\\
    I & I & I & Z & I & I & I & I & X & X\\
    I & I & I & I & I & X & I & Y & X & X\\
    I & I & I & I & I & Z & I & X & I & X\\
    I & I & I & I & I & I & X & X & X & X\\
    I & I & I & I & I & I & Z & Z & I & X\\
    \hline\hline
    I & I & I & X & I & I & I & Z & X & Y\\
    I & I & I & I & I & I & I & Y & Y & Z
  \end{array}\right)\label{eq:gen_10_3_2_3}
  \end{alignat}
\end{footnotesize}
    \end{minipage}\hfill
    \begin{minipage}{0.35\textwidth}
    \begin{footnotesize}
  \begin{alignat}{5}\arraycolsep0.5\arraycolsep
    \left(\begin{array}{*{11}{c}}
      X & X & I & I & I & I & Z & Z & X & I & Z\\
      Z & I & I & I & I & I & I & I & I & I & X\\
      I & Z & I & I & I & I & I & I & I & I & X\\
      I & I & X & I & I & Z & I & X & Z & I & I\\
      I & I & Z & I & I & I & I & I & X & I & I\\
      I & I & I & X & I & Z & Y & Z & X & Y & X\\
      I & I & I & Z & I & I & I & I & I & X & I\\
      I & I & I & I & X & Z & Z & I & I & X & I\\
      I & I & I & I & Z & Z & X & X & I & I & I\\
      I & I & I & I & I & Y & X & Y & I & X & I\\
      \hline
      I & I & I & I & I & Z & I & X & I & X & X\\
      I & I & I & I & I & I & Z & Z & X & X & I\\
      \hline\hline
      I & X & I & I & I & I & I & X & Y & Z & I\\
      I & I & I & I & I & I & X & I & X & Y & Z
    \end{array}\right)\label{eq:gen_11_1_2_4}
  \end{alignat}
\end{footnotesize}
    \end{minipage}\hfill \
    \medskip
\end{table*}

\begin{footnotesize}
  \begin{alignat}{5}\arraycolsep0.5\arraycolsep
    \left(\begin{array}{*{7}{c}}
      X & I & I & Z & Y & Y & Z\\
      Z & I & I & I & I & I & X\\
      I & X & I & X & Z & I & I\\
      I & Z & I & Z & I & X & X\\
      I & I & X & X & I & Z & I\\
      I & I & Z & Z & X & I & X\\
      \hline
      I & I & I & X & Z & Z & X\\
      I & I & I & Z & X & X & I\\
      \hline\hline
      I & I & I & I & X & Y & Y
    \end{array}\right)\label{eq:gen_7_1_1_3}
  \end{alignat}
\end{footnotesize}%

\noindent The weight enumerators of the associated classical codes are
as follows:

\begin{footnotesize}
\begin{alignat}{5}
\mathcal{W}_{C_0}(X,Y)&{}= X^7+X^5Y^2+2X^4Y^3+7X^3Y^4\nonumber\\&{}\quad+24X^2Y^5+23XY^6+6Y^7\\
\mathcal{W}_{C_0^*}(X,Y)&{}= X^7+X^5Y^2+20X^4Y^3+43X^3Y^4\nonumber\\&{}\quad+72X^2Y^5+83XY^6+36Y^7\\
\mathcal{W}_{C^*}(X,Y){}&= X^7+X^5Y^2+36X^4Y^3+91X^3Y^4\nonumber\\&{}\quad+152X^2Y^5+163XY^6+68Y^7
\end{alignat}
\end{footnotesize}%

It can be seen that all codes contain a single word of weight two, and
hence the minimum distance of the hybrid code is three.

We have not found a hybrid quantum code with parameters
$[\![7,1{:}2,3]\!]_2$ which is not ruled out by linear programming.

For eight qubits, there is a quantum code with parameters
$[\![8,3,3]\!]_2$. Using Lemma~\ref{lemma:quantum_to_classical}, we
obtain an optimal hybrid code with parameters $[\![8,2{:}1,3]\!]_2$, as
well as a code $[\![8,1{:}2,3]\!]_2$.  We have not found a hybrid code
with parameters $[\![8,1{:}3,3]\!]_2$ that might exist.

For nine qubits, we found a hybrid code $[\![9,2{:}2,3]\!]_2$ given in
\eqref{eq;gen_9_2_2_3}. The rows above the single horizontal line
generate the stabilizer of an impure code $[\![9,2,3]\!]_2$.  Taking
all possible products of the two generators below the double
horizontal line in \eqref{eq;gen_9_2_2_3} we obtain the four
translation operators $t^{(1)}=id$, $t^{(2)}$, $t^{(3)}$, and
$t^{(4)}=t^{(2)}t^{(3)}$ used to encode two extra classical bits.

The corresponding weight enumerators are as follows:

\begin{footnotesize}
\begin{alignat}{5}
\mathcal{W}_{C_0}(X,Y)&{}= X^9+2X^7Y^2+8X^5Y^4+4X^4Y^5\nonumber\\&{}\quad+22X^3Y^6+56X^2Y^7+31XY^8+4Y^9\\
\mathcal{W}_{C_0^*}(X,Y)&{}X^9+2X^7Y^2+38X^6Y^3+84X^5Y^4+222X^4Y^5\nonumber\\&{}\quad+494X^3Y^6+562X^2Y^7+443XY^8+202Y^9\\
\mathcal{W}_{C^*}(X,Y){}&= X^9+2X^7Y^2+86X^6Y^3+324X^5Y^4+926X^4Y^5\nonumber\\&{}\quad+1934X^3Y^6+2466X^2Y^7+1835XY^8+618Y^9
\end{alignat}
\end{footnotesize}%

A hybrid code $[\![10,3{:}2,3]\!]_2$ with ten qubits is given in
\eqref{eq:gen_10_3_2_3}. and the corresponding weight enumerators are
given in \eqref{eq:we_C0_10_3_2_3}--\eqref{eq:we_C*_10_3_2_3}.

\begin{footnotesize}
  \begin{alignat}{5}
    \mathcal{W}_{C_0}(X,Y)&{}=X^{10}+3X^8Y^2+6X^6Y^4+10X^4Y^6\nonumber\\&{}\quad+105X^2Y^8+3Y^{10}\label{eq:we_C0_10_3_2_3}\\
    \mathcal{W}_{C_0^*}(X,Y)&{}=X^{10}+3X^8Y^2+80X^7Y^3+186X^6Y^4\nonumber\\&{}\quad+432X^5Y^5+1430X^4Y^6+1584X^3Y^7\nonumber\\
       &{}\quad+2325X^2Y^8+1488XY^9+663Y^{10}\label{eq:we_C0*_10_3_2_3}\\
    \mathcal{W}_{C^*}(X,Y){}&=X^{10}+3X^8Y^2+128X^7Y^3+522X^6Y^4\nonumber\\&{}\quad+1824X^5Y^5+5030X^4Y^6+7872X^3Y^7\nonumber\\
     &{}\quad+9477X^2Y^8+6048XY^9+1863Y^{10}\label{eq:we_C*_10_3_2_3}
  \end{alignat}
\end{footnotesize}%

Via linear programming it is found that this code is optimal in the
sense that it encodes the maximal possible number $m$ of additional
classical bits among all codes $[\![10,3{:}m,3]\!]_2$.

The first non-trivial hybrid code with distance $d=4$ has been found
for eleven qubits. A hybrid code $[\![11,1{:}2,4]\!]_2$ is given in
\eqref{eq:gen_11_1_2_4}. We found a hybrid code $[\![11,4{:}2,3]\!]_2$
as well which is given in \eqref{eq:gen_11_4_2_3}.

\begin{footnotesize}
  \begin{alignat}{5}\arraycolsep0.5\arraycolsep
    \left(\begin{array}{*{11}{c}}
      X & X & I & X & X & Y & Y & Z & Y & I & Y\\
      Z & I & I & I & I & I & I & I & X & I & I\\
      I & Z & I & I & I & I & I & I & X & I & I\\
      I & I & X & I & X & Z & I & Z & I & X & X\\
      I & I & Z & X & I & I & Z & X & I & Y & Y\\
      I & I & I & Z & X & X & Z & X & I & X & I\\
      I & I & I & I & Z & Z & Y & X & I & Y & Z\\
      \hline
      I & I & I & X & I & I & I & Z & I & X & I\\
      I & I & I & I & X & I & I & Z & I & Z & Y\\
      I & I & I & I & I & X & I & I & I & X & Z\\
      I & I & I & I & I & Z & I & Z & I & I & X\\
      I & I & I & I & I & I & X & Z & I & X & X\\
      I & I & I & I & I & I & Z & Z & I & Y & Z\\
      I & I & I & I & I & I & I & Y & I & Y & X\\
      I & I & I & I & I & I & I & I & X & X & X\\
      \hline\hline
      I & X & I & I & I & I & I & I & Z & Y & Y\\
      I & I & I & I & I & I & I & Z & Z & X & Z\\
    \end{array}\right)\label{eq:gen_11_4_2_3}
  \end{alignat}
\end{footnotesize}%

Appending two qubits in the state $\ket{0}$ to the impure quantum code
$[\![11,1,4]\!]_2$ given above the double horizontal line in
\eqref{eq:gen_11_1_2_4}, one obtains an impure code
$[\![13,1,4]\!]_2$.  This code can additionally transmit four
classical bits, i.\,e., one obtains the hybrid code
$[\![13,1{:}4,4]\!]_2$ given in \eqref{eq:gen_13_1_4_4}.

\begin{footnotesize}
  \begin{alignat}{5}\arraycolsep0.5\arraycolsep
    \left(\begin{array}{*{11}{c}:cc}
      \multicolumn{11}{c:}{\begin{minipage}{0.2\hsize}
          \vskip1\baselineskip\raggedright
          the stabilizer part of \eqref{eq:gen_11_1_2_4}
          \vskip0.5\baselineskip\ {}
        \end{minipage}} & I & I\\
      \hdashline
      I & I & I & I & I & I & I & I & I & I & I & Z & I\\
      I & I & I & I & I & I & I & I & I & I & I & I & Z\\
      \hline
      I & I & I & I & I & Z & I & X & I & X & X & I & I\\
      I & I & I & I & I & I & Z & Z & X & X & I & I & I\\
      \hline\hline
      I & X & I & I & I & I & I & I & X & Y & X & X & X\\
      I & I & I & I & I & I & X & I & X & I & I & X & X\\
      I & I & I & I & I & I & I & X & Y & X & Y & X & X\\
      I & I & I & I & I & I & I & X & I & Y & Y & X & I
    \end{array}\right)\label{eq:gen_13_1_4_4}
  \end{alignat}
\end{footnotesize}%

A related construction is the following:
\begin{theorem}\label{theorem:ConstructionX}
Let $\mathcal{C}_1=[\![n,k_1,d_1]\!]_q\subset
\mathcal{C}_2=[\![n,k_2,d_2]\!]_q$ be nested quantum codes.  Further,
let $C_3 =[n_3,k_2-k_1,d_3]_q$ be a classical linear code.  Then there
is a hybrid quantum code
$\mathcal{C}=[\![n+n_3,k_1{:}(k_2-k_1),d]\!]_q$ with $d\ge
\min(d_1,d_2+d_3)$.
\end{theorem}
\begin{IEEEproof}
Let $G_1$ be a generator matrix for the normalizer of $\mathcal{C}_1$,
and let $G_{12}$ together with $G_1$ be a generator matrix for the
normalizer of $\mathcal{C}_2$.  Further, let $G_3$ be a generator
matrix of $C_3$, and let $\omega\in\F_{q^2}\setminus\F_q$. The hybrid
code is given by the following matrix:
\begin{alignat}{5}\def\arraystretch{1.1}
  \left(\begin{array}{c:c}
    0   & \omega I\\
    \hdashline
    G_1 &  0\\
    \hline\hline
    G_{12} & G_3
  \end{array}\right)
\end{alignat}
The matrix above the double horizontal line corresponds to the
normalizer of the impure quantum code obtained by appending $n_3$ qudits
in the state $\ket{0}$ to the code $\mathcal{C}_1$. The distance of
this code is $d_1$. Any vector involving the matrix $G_{12}$ will have
weight at least $d_2+d_3$.  Hence, $d\ge \min(d_1,d_2+d_3)$.
\end{IEEEproof}
From the nested stabilizer codes
$[\![11,1,5]\!]_2\subset[\![11,4,3]\!]_2$ and classical codes
$[n_3,n_3-1,2]_2$, one obtains hybrid codes $[\![13,1{:}1,5]\!]_2$,
$[\![14,1{:}2,5]\!]_2$, and $[\![15,1{:}3,5]\!]_2$.  Similarly, we
have the results shown in Table \ref{tab:ConstructionX}. In the first
column we list the nested quantum codes, in the second column the
parameters of the hybrid codes obtained using Theorem
\ref{theorem:ConstructionX}, and in the last column we give the
parameters of the best known stabilizer code from
\cite{Grassl:codetables} which has the same length and minimum
distance as the hybrid code.

\begin{table}
  \caption{Parameters of hybrid codes obtained from nested quantum
    codes using Theorem \ref{theorem:ConstructionX}.}
  \label{tab:ConstructionX}
\begin{align*}\def\arraystretch{1.1}
  \begin{array}{c|c|c}
    \text{nested codes} & \text{hybrid codes} & \text{largest QECC}\\
    \hline
        [\![17,9,4]\!]_2\subset[\![17,13,2]\!]_2
        &[\![19,9{:}1,4]\!]_2 & [\![19,9,4]\!]_2\\
        &[\![20,9{:}2,4]\!]_2 & [\![20,10,4]\!]_2\\
        &[\![21,9{:}3,4]\!]_2 & [\![21,11,4]\!]_2\\
        &[\![22,9{:}4,4]\!]_2 & [\![21,12,4]\!]_2\\
    \hline
        [\![18,6,5]\!]_2\subset[\![18,10,3]\!]_2
        &[\![20,6{:}1,5]\!]_2 & [\![20,6,5]\!]_2\\
        &[\![21,6{:}2,5]\!]_2 & [\![21,7,5]\!]_2\\
        &[\![22,6{:}3,5]\!]_2 & [\![22,8,5]\!]_2\\
        &[\![23,6{:}4,5]\!]_2 & [\![23,8,5]\!]_2\\
    \hline
        [\![16,2,6]\!]_2\subset[\![16,6,4]\!]_2
        &[\![18,2{:}1,6]\!]_2 & [\![18,2,6]\!]_2\\
        &[\![19,2{:}2,6]\!]_2 & [\![19,2,6]\!]_2\\
        &[\![20,2{:}3,6]\!]_2 & [\![20,4,6]\!]_2\\
        &[\![21,2{:}4,6]\!]_2 & [\![21,5,6]\!]_2\\
    \hline
        [\![28,12,6]\!]_2\subset[\![26,16,4]\!]_2
        &[\![30,12{:}1,6]\!]_2 & [\![30,12,6]\!]_2\\
        &[\![31,12{:}2,6]\!]_2 & [\![31,12,6]\!]_2\\
    \hline
        [\![32,16,6]\!]_2\subset[\![32,21,4]\!]_2
        &[\![34,16{:}1,6]\!]_2 & [\![34,16,6]\!]_2\\
        &[\![35,16{:}2,6]\!]_2 & [\![35,16,6]\!]_2\\
        &[\![36,16{:}3,6]\!]_2 & [\![36,16,6]\!]_2\\
        &[\![37,16{:}4,6]\!]_2 & [\![37,16,6]\!]_2\\
        &[\![38,16{:}5,6]\!]_2 & [\![38,18,6]\!]_2\\
    \hline
        [\![25,5,7]\!]_2\subset[\![25,9,5]\!]_2
        &[\![27,5{:}1,7]\!]_2 & [\![27,5,7]\!]_2\\
        &[\![28,5{:}2,7]\!]_2 & [\![28,5,7]\!]_2\\
        &[\![29,5{:}3,7]\!]_2 & [\![29,6,7]\!]_2\\
        &[\![30,5{:}4,7]\!]_2 & [\![30,8,7]\!]_2\\
    \hline
        [\![27,3,9]\!]_2\subset[\![27,7,6]\!]_2
        &[\![30,3{:}1,9]\!]_2 & [\![30,3,9]\!]_2\\
        &[\![32,3{:}2,9]\!]_2 & [\![32,3,9]\!]_2\\
        &[\![33,3{:}3,9]\!]_2 & [\![33,3,9]\!]_2\\
        &[\![34,3{:}4,9]\!]_2 & [\![34,3,9]\!]_2
  \end{array}
\end{align*}
\end{table}

\section{Discussion}
We have characterized hybrid quantum codes for the simultaneous
transmission of quantum and classical information in terms of
generalized Knill-Laflamme conditions.  Using the framework of CWS
codes/union stabilizer codes, we have formulated a linear program to
obtain bounds on the parameters of codes.  Moreover, we found several
examples of hybrid codes that demonstrate the advantage of
simultaneous transmission of quantum and classical information.

The code conditions derived in Section \ref{sec:code_conditions}
suggest that one should start with good impure quantum codes. Theorem
\ref{theorem:ConstructionX} uses trivial impure codes. In order to
find a direct construction of hybrid codes with good parameters, a
first step could be to develop methods to construct good non-trivial
impure codes.

\section*{Acknowledgments}
The authors would like to thank Fred Ezerman, Martin R\"otteler, and
Hui Khoon Ng for fruitful discussions. BZ is supported by NSERC and
CIFAR.



\end{document}